\documentclass[letterpaper,titlepage,11pt]{article}
\pdfoutput=1

\usepackage[T1]{fontenc} 
\usepackage{hyperref}
\usepackage{amssymb,amsmath,amsfonts}
\usepackage{epsfig}
\usepackage{array}
\usepackage{booktabs}
\usepackage{subcaption}
\usepackage[dvipsnames]{xcolor}
\usepackage{siunitx}
\DeclareSIUnit\gauss{G}
\DeclareSIUnit\erg{erg}

\usepackage{tikz}
\usetikzlibrary{calc,through,backgrounds,patterns,trees,positioning,arrows,shapes,chains,%
decorations.pathmorphing,decorations.markings,decorations.pathreplacing,fadings}

\newcommand{\CF}{\mathcal{F}}

\newcommand{\CH}{\mathcal{H}}
\newcommand{\CL}{\mathcal{L}}
\newcommand{\CO}{\mathcal{O}}
\newcommand{\CN}{\mathcal{N}}

\newcommand{\CJ}{\mathcal{J}}

\newcommand{\beq}{\begin{equation}}
\newcommand{\eeq}{\end{equation}}

\newcommand{\huno}{\hspace{1pt}}
\newcommand{\hdue}{\hspace{2pt}}

\newcommand{\hset}{\hspace{7pt}}

\def\ie{\textit{i.e.} }

\def\aD5{$\overline{\mathrm{D}5}$}

\numberwithin{equation}{section}

\begin{document}

\begin{titlepage}
\ \ \vskip 1.8cm
\begin{center}
 {\huge \bf Phase structure of a holographic}
 \vskip 0.3 cm
 {\huge \bf  double monolayer Dirac semimetal} 
\end{center}
\vskip 1.4cm

\centerline{\large {\bf Gianluca Grignani$\,^{1}$},  {\bf Andrea Marini$\,^{1}$},}
\vskip 0.2cm \centerline{\large  {\bf Adriano-Costantino Pigna$\,^{1}$}
and
{\bf Gordon W. Semenoff$\,^{2}$} }

\vskip 0.7cm

\begin{center}
\sl $^1$ Dipartimento di Fisica, Universit\`a di Perugia,\\
I.N.F.N. Sezione di Perugia,\\
Via Pascoli, I-06123 Perugia, Italy
\vskip 0.3cm
\sl $^2$ Department of Physics and Astronomy \\ University of British Columbia \\Vancouver, BC Canada V6T 1Z1
\end{center}
\vskip 0.3cm

\centerline{\small\tt grignani@pg.infn.it, marini@pg.infn.it, }
\centerline{\small\tt pigna@pg.infn.it, gordonws@phas.ubc.ca}

\vskip 1.2cm \centerline{\bf Abstract} \vskip 0.2cm \noindent
We study a holographic D3/probe-D5-brane model of a double monolayer Dirac semimetal in a magnetic field and in the 
	presence of a nonzero temperature. Intra- and inter-layer exciton condensates can form by varying the balanced charge density on the layers, the spatial separation and the temperature. Constant temperature phase diagrams for a wide range of layer
	separations and charge densities are found. The presence of a finite temperature makes the phase diagrams extremely rich and in particular leads to the appearance of a symmetric phase which 
	was missing at zero temperature.

\end{titlepage}



\section{Introduction and summary}
\label{intro}

Double monolayer graphene is a two-dimensional  electronic system formed by two parallel layers of graphene
brought in close proximity 
(on the order of nanometers) but still separated by an insulator so that direct 
transfer of electric charge carriers between the layers can be neglected. 
The attractive Coulomb interaction between electrons and holes can lead to the formation of electron-hole bound states, called excitons.
In a double monolayer system there are two types of excitons: An \emph{inter-layer} exciton which is a bound 
state of an electron in one layer and a hole in the other layer; or an \emph{intra-layer} exciton which is a bound state
an electron and a hole in the same monolayer. Both of these quasi-particles are bosons and they can Bose condense
to form a condensate which we shall call an \emph{inter-layer} condensate and an \emph{intra-layer} condensate,
respectively. 
The possible existence of these exciton condensates is particularly intriguing   because of the possible outstanding
technological applications they can be used for
\cite{4729616,High11072008,2013NatCo...4E1778B,Kuznetsova:10,PhysRevLett.104.027004,PhysRevB.84.184528}. 
This justifies the exceptional scientific interest and the amount of efforts spent in trying to fully understand the 
condensation mechanisms in such double-layer structures
\cite{Gamucci:2014hca,2008JETPL..87...55L,2012NatPh...8..896G,2011PhRvB..83p1401K,2012Natur.488..481N,2008PhRvB..77w3405Z}.

A quantitative analysis of this kind of systems is seriously hampered because of the strength of the Coulomb interaction, 
which prevents a perturbative approach. In this paper we use a non-perturbative holographic model of 
strongly coupled layer systems.
Physical parameters such as external magnetic field, charge density and finite temperatures can be incorporated 
into the solution giving shape to a rich phase diagram with intra- and inter-layer exciton condensate domains.
  
In our model the double monolayer system is described by two parallel, planar 2+1-dimensional defects separated by a length $L$
embedded in 3+1-dimensional Minkowski spacetime. A $U(1)$ potential is turned on to include a balanced charge density on the layers. 
We further switch on a constant external magnetic field and we put the system at finite temperature and study the 
layers in a thermal state. 
The case with unpaired charges is not taken in consideration in this paper. A detailed treatment where the emergent $SU(4)$ symmetry of graphene \cite{Semenoff:1984dq} is accounted can be found in \cite{Grignani:2014vaa}. 

\begin{figure}[!ht]
\begin{center}
		\begin{tikzpicture}[scale=.9]
			\def\lungDt{9};
			\def\largDt{2};
			\def\altDcu{1};
			\def\altDcd{2.5};
			\coordinate (D3a) at (-.5*\lungDt,0);
			\coordinate (D5a) at (-2,0);
			\coordinate (D5b) at (2,0);
			\fill[Cerulean!70,opacity=.8,path fading=south] (D5a) --  ++(2,2)  -- ++(0,-\altDcd) -- ++(-2,-2) -- cycle; 
			\fill[Violet!50,opacity=.8,path fading=south] (D5b) --  ++(2,2)  -- ++(0,-\altDcd) -- ++(-2,-2) -- cycle; 
			\fill[Green!50,opacity=.8] (D3a) --  ++(\lungDt,0)  -- ++(2,\largDt) -- ++(-\lungDt,0) -- cycle; 
			\draw [<->,thick,shorten <= 2, shorten >=2] (-1,1) -- + (4,0) node[midway, fill= Green!40, inner sep=2pt] {\scriptsize $L$};
			\fill[Cerulean!70,opacity=.8] (D5a) --  ++(0,\altDcu)  -- ++(2,2) -- ++(0,-\altDcu) -- cycle; 
			\fill[Violet!50,opacity=.8] (D5b) --  ++(0,\altDcu)  -- ++(2,2) -- ++(0,-\altDcu) -- cycle; 
			\draw[Green!60!black,line width=2](D3a) node[above right=0pt and 5pt] {\footnotesize D3} --  ++(\lungDt,0)  -- ++(2,\largDt);
			\draw[Cerulean!60!black,line width=1.5] ($(D5a)-(0,\altDcd)$) --  ++(0,\altDcu+\altDcd) node[midway,anchor=north west]{\footnotesize D5} -- ++(2,2);
			\draw[Violet!80!black,line width=1.5] ($(D5b)-(0,\altDcd)$) --  ++(0,\altDcu+\altDcd) node[midway,anchor=north west]{\footnotesize $\overline{\mbox{D5}}$} -- ++(2,2);
			\draw[Blue,line width=2,line cap=round] (D5a) -- +(2,2) node[near end] (A) {};
			\draw[Violet,line width=2,line cap=round] (D5b) -- +(2,2)  node[near end] (B) {};
			\fill[white,path fading=north] (-3,-3) rectangle +(8,2.5);
			\node (T) at (1.8,3) {\footnotesize $2+1$-dim defects};
			\draw [blue!70!black,line width=1,->] (A) to[in=-120,out=0] (T); 
			\draw [blue!70!black,line width=1,->] (B) to[in=-60,out=180] (T); 
		\end{tikzpicture}
		\end{center}	
		\caption{\small Scheme of the D3/D5-\aD5 -brane system configuration.}	
		\label{fig:D3D5conf}
\end{figure}
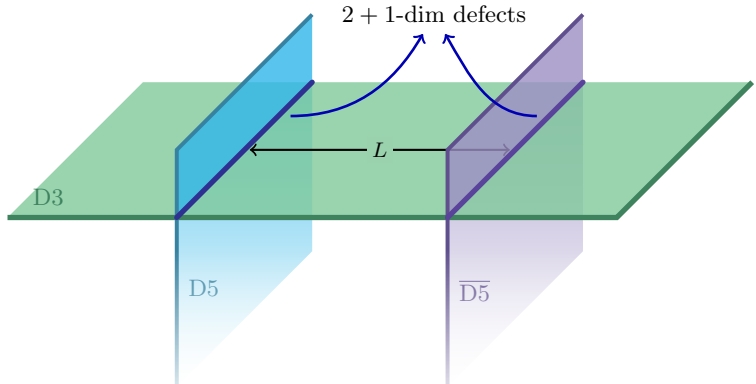

The holographic top-down model we employ for the relativistic defect quantum field theory setup described 
above is given by a D3/D5-\aD5-brane system. This same model was studied before in \cite{Grignani:2014vaa} but only 
at zero temperature. Here we analyze the finite temperature generalization of this holographic system.
We consider the background generated by a stack of $N$ coincident non-extremal D3-branes. In the near horizon limit this gives rise to the well-known $AdS_5\times S^5$ black hole 
geometry. Then we embed the D5 and the \aD5 (anti-D5) branes as probes of this background
in such a way that they have a 2+1-dimensional intersections with the 3+1-dimesional D3-brane worldvolume, 
as depicted in Figure~\ref{fig:D3D5conf}. 
These intersections are the dual holographic realizations of the Dirac semimetal monolayers. 
We use a brane-anti-brane pair because, as we will discuss, they can partially annihilate and the annihilation will be the string theory dual of the formation of an inter-layer exciton condensate.

According to AdS/CFT correspondence we can use this brane setup to describe two 2+1-dimensional defect theories interacting 
through the exchange of SU($N$) $\CN=4$ SYM degrees of freedom that propagate in 3+1-dimensions.
As usual the string model is effective in the strong coupling regime, \ie when the 't Hooft coupling $\lambda\equiv g_s N/4\pi=g_{YM}^2 N$
is large. Here, $g_s$ is the string couplig and $g_{YM}$ is the coupling constant of the gauge fields in the defect quantum field theory. 
However this is not enough and in order for the dual theory to be tractable we have to take $N\rightarrow \infty$, while keeping $\lambda\equiv g_s N/4\pi=g_{YM}^2 N$ constant, in such a way to suppress quantum effects.

In practice the procedure we have to follow in order to study the holographic model consists in
finding all the allowed configurations for the D5- and the \aD5-branes and then comparing
their free energy in order to determine the favored one. In the probe regime this is achieved by looking for the solutions of
the Dirac-Born-Infeld action for the D5 and the \aD5 in the $AdS_5\times S^5$ black hole background 
with the suitable boundary conditions.\footnote{The D5-branes we are considering are probes in
	a finite temperature background.
	Using the Dirac-Born-Infeld action to describe the D5-branes dynamics we are neglecting the 
	thermal degrees of freedom induced on the D5's by the background. 
	However this is a good approximation in the large $N$ limit. In order to take into account the
	thermal excitation of the probes one could use the alternative approach developed 
	in~\cite{Grignani:2010xm,Grignani:2011mr} based on the blackfold approach \cite{Emparan:2009at}.} 
The geometry of the brane worldvolume
is then related through the AdS/CFT dictionary to the relevant field theory quantities, in this case the expectation values of the 
condensates. 
The geometry of the D5 and the \aD5 at the boundary is fixed to be $AdS_4 \times S^2$, with maximal $S^2$ radius.
As the branes penetrate the $AdS_5$ bulk their geometry can change in two ways: The radius of the $S^2$ can decrease and 
possibly also become zero before reaching the horizon. 
This would correspond to the presence of a non-zero value for the intra-layer condensate.
The other possible modification of the geometry is induced by the presence of the pair of branes: Being like a particle-hole pair, 
the D5-brane and the \aD5-brane have a tendency to annihilate. Note though that we impose boundary conditions which prevent 
their complete annihilation. As the D5-brane approaches the boundary of the $AdS_5$ we require that it has to be parallel 
to the \aD5-brane and it has to be separated from the \aD5-brane by a spatial distance $L$. 
However, as they enter into the bulk of $AdS_5$, they can still partially annihilate by joining smoothly at some finite value of the
radial coordinate of the $AdS_5$ space. 
This joining of the brane and anti-brane is the AdS/CFT dual of inter-layer exciton condensation.  

\bigskip

Before considering the case we are intrested in, namely the double monolayer system, 
let us briefly review the results that the D3/D5 model provides in the simplest case of a single monolayer
\cite{Evans:2010hi,Kristjansen:2012ny,Kristjansen:2013hma}. Of course in this case only the intra-layer condensation can occur.
At zero temperature and in absence of magnetic fields and charge densities, 
the D5-brane configuration is supersymmetric and conformally invariant. 
The D5 worldvolume has an $AdS_4\times S^2$ geometry and stretches from the boundary of $AdS_5$ to the Poincar\'e horizon.
This is a maximally symmetric solution that corresponds to a configuration without any condensate.  
Its quantum field theory dual is well known \cite{Karch:2000gx,Karch:2000ct,DeWolfe:2001pq,Erdmenger:2002ex}. 

As soon as one introduces an external magnetic field on the brane world volume, the single D5-brane geometry changes
drastically \cite{Filev:2009xp}. The brane pinches off and truncates at a finite $AdS_5$ radius, 
before it reaches the horizon and then it has a ``Minkowski embedding''. This has to be interpreted as 
formation of a fermion anti-fermion condensate which induces a mass gap, since the open
strings stretching from the D5-brane to the horizon have a minimum length greater than zero.

When a nonzero charge density is introduced in addition to the magnetic field, the D5-brane embedding switches 
to a ``Black Hole embedding'', meaning that the brane stretches from the boundary of $AdS_5$ to the event horizon. 
This is due to the fact that the brane carries worldvolume electric flux. It is not possible for the worldvolume to pinch off 
smoothly unless there is a sink to absorb the electric flux. Such a sink is provided by a density of fundamental strings suspended between the worldvolume and the horizon. However, fundamental strings with the necessary configuration always have a larger tension than the D5-brane, and they would pull the D5-brane to the horizon \cite{Kobayashi:2006sb}. 
This means that the theory no longer has a mass gap. 

Moreover for high enough chemical potential the system undergoes a phase transition with BKT scaling to a phase in which 
the condensate disappears and the symmetry is restored \cite{Jensen:2010ga}. 
Also the temperature promotes the symmetry restoration but at finite temperature the transition is no longer of BKT type.
The full phase diagram for the single monolayer case can be found in \cite{Evans:2010hi}.



D3/D5-\aD5-brane model for the double monolayer system has been studied at zero temperature in \cite{Evans:2013jma,Grignani:2014vaa}.  
In presence of a magnetic field $B$ and keeping a vanishing charge density on the branes 
the phase structure is determined by the competition of two opposite effects: On the one hand the magnetic 
field induces the pinching off of the brane worldvolume, which corresponds
to the the condensation in the intra-layer channel; on the other hand the proximity of the branes 
promotes the partial annihilation of the brane and the anti-brane, which is the holographic dual of 
the inter-layer condensation. These two effect are governed by the (dimensionless) parameter 
$\sqrt{B} L$. Exploring all the possible values for the latter one sees that the phase diagram 
contains only two regions: For small $\sqrt{B} L$ the stable phase is the one with the inter-layer 
condensate only; increasing $\sqrt{B} L$ the system undergoes a phase transition to a state in which only 
the intra-layer condensate is present \cite{Evans:2013jma}. 

Turning on a balanced charge density $q$ on each brane introduces a new dimensionless parameter to play with,
 $q/\sqrt{B}$ (or $\mu/\sqrt{B}$, $\mu$ being the chemical potential). 
In this case a new region appears in the phase diagram, which corresponds to a configuration where 
both the inter- and intra-layer condensates are simultaneously present \cite{Grignani:2014vaa}. 
Similar results can be obtained using top-down model with D7-brane probes instead of 
D5's \cite{Grignani:2014tfa,Grignani:2012qz}.

In \cite{Filev:2013vka,Filev:2014bna} a different model of flavor chiral symmetry breaking in a (2+1)-dimensional defect gauge theory of 
strongly coupled fermions was proposed by introducing probe D5/anti-D5-flavor branes on the Klebanov-Witten background. Introducing finite temperature, an $r$-dependent profile in the $x_3$-direction transverse to the (domain wall) defects and 
an external magnetic field the thermodynamics of the resulting configurations was studied and a detailed phase diagram of the model was established.

\begin{figure}[!ht]
\centerline{\includegraphics[scale=0.45]{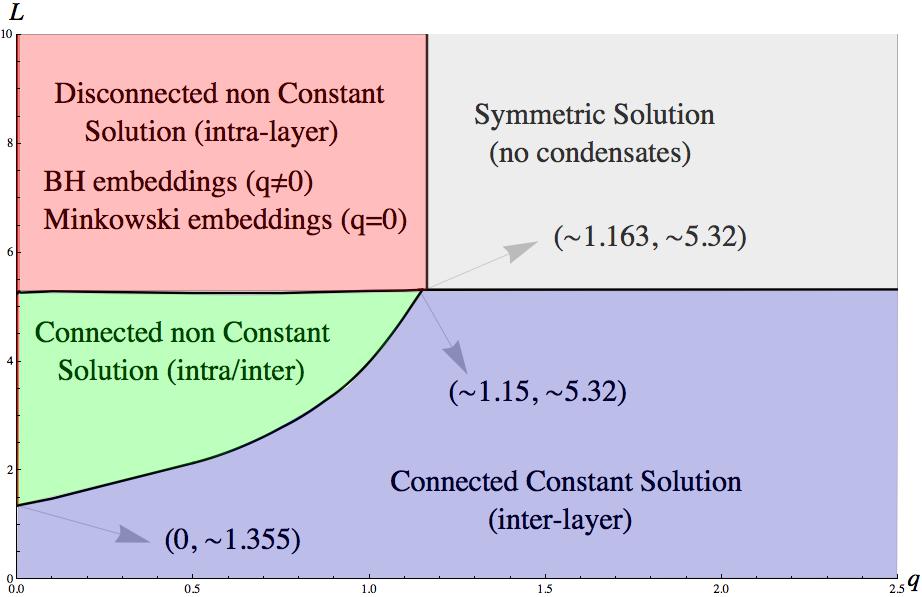}}
\caption{\small Slice of the phase diagram for D3/D5-\aD5-brane model with a magnetic field and 
balanced charge densities, at a fixed value of the temperature, $T = 0.1 \times \frac{\sqrt{2}}{\pi}$.
The brane separation is plotted on the vertical axis and the charge density on the horizontal one.
We employ units where the length scale $\frac{\lambda^{1/4}}{\sqrt{2\pi \, B}}$ is set to one.
}
\label{fig:phaseintro}
\end{figure}

In this paper we further generalize the model considered in \cite{Grignani:2014vaa}
by taking into account also the effect of the temperature: Namely we study the D3/D5-\aD5-brane 
model with a magnetic field, balanced charge densities on the branes and at finite temperature. 
Remarkably the temperature makes the phase structure of the system even richer and unlike before all the four possible 
phases play a dominant role for suitable domains in the phase space.
We will show indeed that, besides the three phases already present at zero temperature, 
for sufficiently high charge densities and separations, a symmetric phase emerges where both 
the intra- and inter-layer condensates are vanishing. 
This effect is completely consistent with the fact that both the charge density and the temperature 
induce a symmetry restoration. 
An example of phase diagram is shown in Figure~\ref{fig:phaseintro} 
for a particular value of the temperature.

\bigskip

In the remainder of the paper, we will describe the quantitative analysis which leads to the phase diagram of the D3-probe-D5-brane system. In Section~\ref{sec:setup} we will discuss the geometry of the D5-branes embedded in $AdS_5\times S^5$ black hole background. Then we will find the equations of motion and we will list all their possible solutions. In Section~\ref{sec:solutions} we will give a more detailed characterization of the solutions, showing the procedures we used to find them. In Section~\ref{sec:free_en} we will show the numerical analysis of the solutions and in particular the comparison of the their free energies as a function of the layer separation, 
the charge density and the temperature. This will allows us to draw of the phase diagrams of the system. 
In Appendix~\ref{sec:cgsunits} we will restore the physical units to give an estimate of the orders of magnitude of the parameters.


\section{Geometric setup}
\label{sec:setup}

We consider a pair of probe branes, a D5 brane and an \aD5  brane suspended in $AdS_{5}\times S^5$ black hole background, with a metric of the form
\begin{equation}\label{eq:ads5s5metric}
	ds^2= r^{2}(-h(r)dt^2 + dx^2 + dy^2 +dz^2) +\frac{dr^2}{r^{2}h(r)} +d\psi^2 +\sin^{2}{\psi} d\Omega^2_{2} + \cos^{2}{\psi}d\tilde{\Omega}^2_{2}
\end{equation}
Here, the coordinates of $S^5$ are a fibration of the $5$-sphere by two $2$-spheres over the
interval $\psi \in [0,\pi/2]$ and $(t, x, y, z, r)$ are coordinates of $AdS_5$ spacetime. 
The metrics of the two $2$-spheres $S^2$ and $\tilde{S}^2$ are $d^2\Omega_2=d\theta^2+\sin^2\theta d\phi^2$ and $d^2\tilde{\Omega}_2=d\tilde{\theta}^2+\sin^2\tilde{\theta} d\tilde{\phi}^2$ respectively.
The $h(r)$ factor appearing in the temporal and radial components is
\begin{equation}
h(r)=1-\frac{r_{h}^{4}}{r^4}
\label{eq:eventhorizon}
\end{equation} 
where $r_h$ is the radius of the event horizon and $T=r_{h}/\pi$ is the Hawking temperature.
Our choice of branes coordinates is shown in Table~\ref{tab:coords1}. 
\begin{table}[!ht]
\begin{center}
\begin{tabular}{ccccccccccc}
\toprule
 & $t$ & $x$ & $y$ & $z$ & $r$ & $\psi$ & $\theta$ & $\phi$ & $\tilde{\theta}$ & $\tilde{\phi}$ \\
\midrule
  D3 & $\bullet$ & $\bullet$ & $\bullet$ & $\bullet$ &  & & & & & \\
  D5/\aD5 & $\bullet$ & $\bullet$ & $\bullet$ & $z(r)$ & $\bullet$ & $\psi(r)$ & $\bullet$ & $\bullet$ & & \\
\bottomrule
\end{tabular}
\end{center}
\caption{\small Choice of the D5- and \aD5-brane embedding coordinates.}
\label{tab:coords1}
\end{table}

We require Poincar\'e invariance in the $2+1$-dimensional intersection of the branes.  So both the D5- and the \aD5 -brane wrap the $(t,x,y)$ subspace of $AdS_5$. We also assume that the D5- and \aD5 -brane worldvolumes wrap the $2$-sphere $S^2$ with coordinates $\theta,\phi$ providing an $SO(3)$ symmetry for the solutions. 
The D5- and \aD5 -branes sit at points in the remaining directions, $z,\psi,\tilde{\theta},\tilde{\phi}$. We require that $z(r)$ and $\psi(r)$ depend on the radial coordinate $r$ in order to have a non trivial dynamics of the embeddings.
Note that the point $\psi=\frac{\pi}{2}$ where $S^2$ is maximal has an additional $SO(3)$ symmetry.

However, in order to find all the possible solutions and to avoid the presence of off-diagonal terms in the metric 
we move to different coordinates. Following the steps of ref. \cite{Evans:2013jma} we define a 
new radial coordinate $w$ in such a way that
\begin{equation}
\frac{dr^2}{h(r) r^2}=\frac{dw^2}{w^2} \hspace{20pt} \rightarrow \hspace{20pt} 
\frac{r dr}{\sqrt{r^4 - r^{4}_{h}}}=\frac{dw}{w} \, .
\end{equation}
By integrating the last relation and requiring that $w=r$ in the zero-temperature limit $r_{h}=0$ we get
\[
w(r)=\frac{\sqrt{r^2+\sqrt{r^4 - r^{4}_{h}}}}{\sqrt{2}}\, . 
\]
The event horizon $r=r_h$ in the new coordinate is placed at $w=w_h\equiv r_h/\sqrt{2}$. 
The inverse coordinate transformation is then 
\[
r^2=\frac{w^4 + w_{h}^{4}}{w^2}\, .
\]
The $AdS_5$ metric components in the new coordinate system turn out to be
\begin{equation}
g_{tt}=h(r) r^2=\frac{(w^4 - w_{h}^{4})^2}{w^2 (w^4 + w_{h}^{4})}, \hspace{15pt} g_{xx}=g_{yy}=g_{zz}=r^2=\frac{w^4 + w_{h}^{4}}{w^2} \, ,
\label{eq:coeff}
\end{equation}
so that we can rewrite the metric \eqref{eq:ads5s5metric} as
\begin{equation}\label{eq:radialmetric}
\begin{split}
	ds^2=  -\frac{(w^4 - w_{h}^{4})^2}{w^2 (w^4 + w_{h}^{4})} dt^2 &+ \frac{w^4 + w_{h}^{4}}{w^2}(dx^2+dy^2+dz^2) \\
	& +\frac{dw^2}{w^2}+d\psi^2 +\sin^{2}{\psi} \, d\Omega_{2} ^2+ \cos^{2}{\psi}\, d\tilde{\Omega}_{2}^2 \, .
\end{split}
\end{equation}
The last step is to switch to planar coordinates by defining 
\[
\rho=w \sin\psi\, , \qquad l=w \cos\psi\, ,
\]
in such a way that $\rho^2+l^2=w^2$.
The $AdS_5 \times S^5$ metric finally becomes
\begin{equation}\label{eq:finalmetric}
\begin{split}
	ds^2=-\frac{(w^4 - w_{h}^{4})^2}{w^2 (w^4 + w_{h}^{4})} dt^2 &+  \frac{w^4 + w_{h}^{4}}{w^2}(dx^2+dy^2+dz^2) \\
	&+ \frac{d\rho^2}{\rho^2+l^2}+ \frac{\rho^2}{\rho^2+l^2}d\Omega_{2}^2+ \frac{dl^2}{\rho^2+l^2}+ \frac{l^2}{\rho^2+l^2}d\tilde{\Omega}_{2}^2.
\end{split}
\end{equation}
Now the $\rho$-dependent dynamic variables are $z(\rho)$ and $l(\rho)$.  
With these new coordinates the point $\psi=\frac{\pi}{2}$ where $S^2$ is maximal becomes the point $l=0$.
The branes coordinates ansatz is shown in Table~\ref{tab:coords2}.
\begin{table}[!ht]
\begin{center}
\begin{tabular}{ccccccccccc}
\toprule
 & $t$ & $x$ & $y$ & $z$ & $\rho$ & $\theta$ & $\phi$ & $l$ & $\tilde{\theta}$ & $\tilde{\phi}$ \\
\midrule
  D3 & $\bullet$ & $\bullet$ & $\bullet$ & $\bullet$ &  & & & & & \\
  D5/\aD5 & $\bullet$ & $\bullet$ & $\bullet$ & $z(\rho)$ & $\bullet$ & $\bullet$ & $\bullet$ & $l(\rho)$ & & \\
\bottomrule
\end{tabular}
\end{center}
\caption{\small Choice of the D5- and \aD5-brane embedding coordinates in the planar frame.}
\label{tab:coords2}
\end{table}

The asymptotic behavior at $\rho\rightarrow\infty$ for the embedding function  $l(\rho)$ is 
\begin{equation}
\label{eq:asymptemb}
l(\rho) \rightarrow c_0 + \frac{c_1}{\rho} + \dots
\end{equation}
while the asymptotic expansion for $z(\rho)$ is such that the D5-brane and \aD5-brane remain separated by a distance $L$ as $\rho\rightarrow\infty$
\begin{equation}
\label{eq:asymptzeta}
z(\rho)\rightarrow \pm \frac{L}{2} \mp \frac{f}{\rho^5}+\dots
\end{equation} 
Every coefficient in the last two formulas has a specific physical interpretation in terms of a field theory dual quantity. 
In particular $c_0$ and $c_1$ in eq.~\eqref{eq:asymptemb} are related to the fermion mass and to the expectation value 
of the intra-layer chiral condensate respectively. Intra-layer means that the condensation occurs between fermionic 
species on a single layer.
In eq.~\eqref{eq:asymptzeta} $f$ is proportional to the expectation value of the inter-layer chiral condensate, meaning that the condensation occurs between a fermionic species on one layer and a fermionic species on the other layer. 
In this paper, we will only consider solutions where $c_0 =0.$ This is the boundary condition that is needed for the Dirac fermions in the defect quantum field theory to be massless.

When $c_1=0$ and $l=0$ for all values of $\rho$ we have a chirally symmetric phase. If $c_1\neq 0$ this implies $l(\rho)\neq 0$. This breaks the maximal $SO(3)$ and we have a phase where the chiral symmetry is spontaneously broken with the formation of an intra-layer condensate.

\subsection{Equations of motion}

The Born-Infeld action for either the D5-brane or the \aD5 -brane is given by
\begin{equation}
S=-\CN_{5}\int d^{6}\sigma \, \sqrt{-\det{(\gamma+2\pi \alpha ^{'} \CF)}}
\label{eq:DBIaction}
\end{equation}
where $\sigma^a$ are the coordinates of the D$5$-brane worldvolume, 
$\gamma_{ab}(\sigma)$ is the induced metric on the D$5$-branes 
\begin{equation}
\begin{split}
	\gamma_{ab} d\sigma^a d\sigma^b=-\frac{(w^4 - w_{h}^{4})^2}{w^2 (w^4 + w_{h}^{4})} dt^2 &+  \frac{w^4 + w_{h}^{4}}{w^2}(dx^2+dy^2) \\ 
	&+ \frac{1+l'{}^2+\left(w^4 + w_{h}^{4}\right)z'{}^2}{w^2}\, d\rho^2+ \frac{\rho^2}{w^2}d\Omega_{2}^2
\end{split}
\end{equation}
and $\CN_5$ is defined as
\[
\CN_5 = \frac{\sqrt{\lambda} N N_5}{2\pi^3} V_{2+1}
\]
where $V_{2+1}$ is the volume of the $2+1$-dimensional space-time, $N$ is the number of D3-branes and $N_5$ is the number of D$5$-branes.
The field strength $2$-form $\CF$, needed to introduce a $U(1)$ charge density on the D5-branes and an external magnetic field, is given by
\begin{equation}
\frac{2\pi}{\sqrt{\lambda}}\CF=a_{0}'(\rho)\hdue d\rho\wedge dt+b\hdue dx\wedge dy.
\end{equation}
In this equation, $b$ will give a constant external magnetic field in the holographic dual and $a_{0}(\rho)$ will result in the world volume electric field related to a non-zero U(1) charge density in the field theory.  The magnetic field $B$ and the gauge field $A_0$ are defined as
\begin{equation}
b=\frac{2\pi}{\sqrt{\lambda}}B\, , \qquad a_0 =\frac{2\pi}{\sqrt{\lambda}}A_0.
\end{equation}
The asymptotic behavior of the gauge field is 
\begin{equation}
a_{0}(\rho)=\mu - \frac{q}{\rho}+\dots
\label{eq:asympgauge}
\end{equation}
where $\mu$ and $q$ are the chemical potential and the charge density, respectively.

The Born-Infeld action for the particular embeddings we are considering is
\begin{equation}
\label{eq:DBIemb}
\begin{split}
	S=\CN_5 \int d\rho\, & \frac{\rho^2\sqrt{(b^2 w^4 + (w^4 +w_{h}^4)^{2}) }}{w^{6} \sqrt{w^4 + w_{h}^4)}} \, \times \\
	&\sqrt{ -w^4 (w^4 +w_{h}^4) a_{0}'^{2}  + (w^4 - w_{h}^4)^2 (1+l'^2 +(w^4 +w_{h}^4) z'^2 )}\, .
\end{split}
\end{equation} 
It is more convenient to switch to magnetic units where the the magnetic field $b$ is rescaled to one by means of the following replacements
\begin{equation}
\label{eq:replacements} 
\begin{split}
	\rho\rightarrow & \sqrt{b}\rho\, , \qquad l\rightarrow \sqrt{b}l\, , \qquad f\rightarrow b^2 f\, , \qquad w \rightarrow \sqrt{b}w, \\
	&q\rightarrow bq\, , \qquad L\rightarrow \sqrt{b}L\, , \qquad \CF_i\rightarrow b^{3/2}\CF_i\, .
\end{split}
\end{equation}
The Born-Infeld action in magnetic units becomes 
\begin{equation}
\label{eq:DBIfinal}
\begin{split}
S=\CN_5 \int d\rho\, & \frac{\rho^2\sqrt{(w^4 + (w^4 +w_{h}^4)^{2}) }}{w^{6} \sqrt{w^4 + w_{h}^4)}} \, \times \\
&\sqrt{ -w^4 (w^4 +w_{h}^4) a_{0}'^{2}  + (w^4 - w_{h}^4)^2 (1+l'^2 +(w^4 +w_{h}^4) z'^2 )}\, .
\end{split}
\end{equation} 

The variational problem of extremizing the Born-Infeld action \eqref{eq:DBIfinal} involves two cyclic variables, $a_0(\rho)$ and $z(\rho)$. Being cyclic, their canonical momenta are constants,
\begin{equation}
\label{eq:conjmomenta}
Q=-\frac{\delta S}{\delta A_{0}'}=\frac{2\pi \CN_5}{\sqrt{\lambda}}q\, , \qquad \Pi_z=\frac{\delta S}{\delta z'}=\CN_5 f.
\end{equation}  
We can invert these relations in order to write $a_0 '$ and $z'$ as function of the parameters $q$ and $f$  finding 
\begin{equation}
\label{eq:qinverse}
a_{0}' (\rho)= \frac{qw^2 (w^4 - w_{h}^4)^2 \sqrt{1+l'^2}}{\sqrt{(w^4 + w_{h}^4) \left(-f^2 w^{12}+(w^4 - w_{h}^4)^2 (q^2 w^8 +(w^4 +(w^4+w_{h}^4)^2)\rho^4) \right) }}
\end{equation}
and
\begin{equation}
\label{eq:finverse}
z' (\rho)= \frac{fw^6 \sqrt{1+l'^2}}{\sqrt{(w^4 + w_{h}^4) \left(-f^2 w^{12}+(w^4 - w_{h}^4)^2 (q^2 w^8 +(w^4 +(w^4+w_{h}^4)^2)\rho^4) \right) }}.
\end{equation}

Using \eqref{eq:qinverse} and \eqref{eq:finverse} we can write the Euler-Lagrange
equation of motion for $l$ in term of the parameters $q$, $f$ and $w_h$. This is a
second order ODE for $l(\rho)$, and being quite a long and complicated equation, we do not show it explicitly. 
This equation admits a trivial constant solution, $l=0$. The other solutions with the right asymptotic
behavior \eqref{eq:asymptemb} have to be found numerically. However an asymptotic expansion of the solution can be worked out 
analytically up to desired order: The first few terms in this expansion, fixing $c_0=0$ in eq.~\eqref{eq:asymptemb}, 
are for instance 
\begin{equation}
\label{eq:asymbeha}
\begin{split}
&l(\rho) =  \frac{c_1}{\rho} + \frac{c_1 (-2+c_{1}^2 -q^2 +2 w_{h}^4)}{10 \, \rho^5} \, +\\
&\frac{c_1 (32+15 c_{1}^4 +30 f^2 +46 q^2 + 15 q^4 +56w_{h}^4 +44 q^2 w_{h}^4 + 72 w_{h}^8 +c_{1}^2 (4-30 q^2 -4 w_{h}^4))}{360\, \rho^9} \\
& + \CO\left(\frac{1}{\rho^{13}}\right) \, .
\end{split}
\end{equation}
We see that this asymptotic expansion is written in terms of the modulus $c_1$ which, as we already pointed out, is
related to the expectation value of the intra-layer condensate.
The constant solution $l=0$ corresponds to a configuration without intra-layer condensate.

\subsection{Classification of solutions}

We can distinguish four types of solutions of the equations of motion.
\begin{itemize}
	\item Unconnected ($f=0$) constant ($c_1 =0$) solutions: They correspond to state of the double monolayer 
	where both the intra- and inter-layer condensates vanish. We will call them ``symmetric'' or ``black'' solutions.
	These solutions reach the event horizon, namely they are Black Hole embedding. 
	\item Connected ($f\neq 0$) constant ($c_1 =0$) solutions:  $z(\rho)$ has a non trivial profile with a boundary condition
	given by eq.~\eqref{eq:asymptzeta}. These solutions correspond to double layers with a non-zero inter-layer 
	condensate and a zero intra-layer condensate. They will be called ``connected constant'' or ``blue'' solutions. 
	\item Unconnected ($f=0$) non-constant ($c_1 \neq 0$) solutions: Non constant means that the embedding 
	function $l$ is $\rho$-dependent. Its asymptotic behavior is given by \eqref{eq:asymbeha}. 
	Instead the $z$ profile is such that the D5-branes are kept apart by a fixed distance $L$ for every value of $\rho$.  
	These embeddings correspond to double monolayers with a non-zero intra-layer condensate
	 and a vanishing inter-layer condensate. 
	We will refer to these as ``unconnected non-constant'' or ``red'' solutions. 
	These solutions can be in principle either Minkowski or Black Hole embeddings depending on the value of the charge density.  
	\item Connected ($f\neq 0$) non-constant ($c_1 \neq 0$) solutions: 
	The profiles along both $z(\rho)$ and $l(\rho)$ are non trivial. 
	These solutions correspond to double monolayers with both intra-layer and inter-layer condensates. 
	These will be the ``connected non-constant'' or ``green'' solutions.
\end{itemize}

Note that the connected configurations are allowed only if the profiles of the D5-brane and the \aD5-brane join smoothly
at a certain $\rho=\rho_t$. This is possible only if $z'(\rho_t)=\infty$, thus $\rho_t$
is the point where the denominator of \eqref{eq:finverse} vanishes. 
We will refer to this point as the ``turning point'' of the solution.
One can construct such configurations only if the whole system is charge neutral, \ie $q\equiv q_{D5}=-q_{\bar{D}5}$.
Furthermore we must also have $f\equiv f_{D5}=-f_{\bar{D}5}$. We will take into account only configurations where these
condition are satisfied.

\subsection{Routhians and brane separation}

Consider the on-shell action \eqref{eq:DBIfinal} evaluated on solutions of the equations of motion 
$\CF_1 \equiv S(l,z,a_0)/\CN_5$. This gives the free energy of our system as a function of 
the chemical potential $\mu$ and the separation $L$. Indeed if we take a variation of these parameters the 
variation of the on-shell action is
\begin{equation}
\delta\CF_1=\int_{0}^{\infty}d\rho \hdue \Bigl(\frac{\partial \CL}{\partial l'}\delta l + \frac{\partial \CL}{\partial a_0'}\delta a_0+ \frac{\partial \CL}{\partial z'}\delta z\Bigr)\huno=\huno -q\delta\mu + f\delta L 
\end{equation} 
where we have used \eqref{eq:conjmomenta} and $\delta l\sim 1/\rho \simeq 0$.

Since $a_0$ and $z$ are two cyclic variables we can write down two Routhians 
related to the on-shell action by Legendre transforms. 
Performing a Legendre transform with respect to the charge density $q$ we can define the Routhian $\CF_2$
\begin{equation}
\label{eq:routh1}
\begin{split}
	&\CF_2 = \huno \CF_1 +q\huno \mu \huno = \huno \CF_1 + \int \huno d\rho \hdue q \hdue  a_{0}' (\rho) \, =\\
	&\int d\rho\, \frac{\left(w^4-w_{h}^4\right)
		\sqrt{\left(q^2 w^8 + \left(\left(w^4 + w_{h}^4\right)^2 + w^4\right) \rho^4\right) \left(1 + l'(\rho)^2
				+ \left(w^4 + w_{h}^4\right) z'(\rho)^2\right)}}{w^6 \sqrt{w^4+w_{h}^4}}.
\end{split}
\end{equation}
where we used 
\[
a_0'=\frac{qw^2 \left(w^4 - w_{h}^4\right)\sqrt{1+l'^2 +\left(w^4 + w_{h}^4\right)z'^2 }}
{\sqrt{\left(w^4 + w_{h}^4\right)\left(q^2 w^8 + \left(w^8 + w_{h}^8 + w^4 \left(1 + 2 w_{h}^4\right)\right) \rho^4 \right)}}. 
\] 
This Routhian is a function of $q$ and $L$ and then it provides the free energy in an ensemble where 
the charge density and the separation are kept fixed.

The last Routhian is obtained performing a second Legendre transform on $\CF_2$
with respect to $L$, giving as a result
\begin{equation}
\label{eq:routh2}
\begin{split}
&\CF_3 = \huno \CF_2 - \int \huno d\rho \hdue f\hdue  z'(\rho) \, = \\
 & \int d\rho\, \frac{\sqrt{\left(-f^2 w^{12} + \left(w^4 - w_{h}^4\right)^2 \left(q^2 w^8 +\left(\left(w^4 + w_{h}^4\right)^2 + w^4\right)  \rho^4\right) \right)
 			\left(1+l'(\rho)^2 \right)} }{w^6\sqrt{w^4 +w_{h}^4}}\, .
\end{split}
\end{equation}
$\CF_3$ is the free energy as a function of $f$ and $q$.

In this paper we work in the ensemble where the charge density $q$ and the separation $L$ are held fixed. Thus we shall
use the Routhian $\CF_2$ to compute the free free energies of the solutions.

The separation of the D$5$ and \aD5 -branes for the connected solutions can be obtained simply 
by integrating the expression \eqref{eq:finverse} for $z'(\rho)$ evaluated on the solutions of the equations of motion
\begin{equation}
\label{eq:separation}
\begin{split}
L&=2\int_{\rho_t}^{\infty} d\rho \hdue z'(\rho)  \\
 &= 2f \int_{\rho_t}^{\infty} d\rho \frac{w^6 \sqrt{1+l'^2}}{\sqrt{(w^4 + w_{h}^4) \left(-f^2 w^{12}+(w^4 - w_{h}^4)^2 (q^2 w^8 +(w^4 +(w^4+w_{h}^4)^2)\rho^4) \right)}}
\end{split}
\end{equation}
where $\rho_t$ is the turning point, \ie the point where the denominator in the integrand vanishes.

\section{Solutions}
\label{sec:solutions}

In the previous section we introduced all the possible types of solutions, now we discuss them in more detail
showing also the procedure we used to find them.

\subsection{Symmetric (black) solutions}

The symmetric configuration is the trivial solution in which both the intra- and inter-layer condensates vanish. 
This means that the D5-branes have flat profiles both along $z$ and $l$ directions. 
Accordingly the separation between the branes is given by $z(\rho)=L$ and the embedding functions are vanishing $l(\rho)=0$. 
The free energy $\CF_2$ of this solution is given by \eqref{eq:routh1} 
where we set $l(\rho)= l'(\rho)=f= 0$ (which in turn implies $w=\rho$) 
\begin{equation}
\label{eq:symmrouth}
\CF_{\mathrm{symm}}\huno=\huno\int d\rho\hdue \Bigl(\frac{\rho^4-w_{h}^4}{\rho^4}\Bigr)\huno\sqrt{\frac{w_{h}^8 + 2\huno w_{h}^4\huno \rho^4 + \rho^4 (1 + q^2 + \rho^4)}{\rho^4 + w_{h}^4}}.
\end{equation}
If we now turn off the temperature by setting $w_{h}=0$, we find $\CF_{\mathrm{symm}}=\int d\rho\hdue\sqrt{1+q^2 +\rho^4}$ 
in agreement with \cite{Grignani:2014vaa}.

Note that the free energy as defined in \eqref{eq:symmrouth} is divergent since as $\rho\rightarrow\infty$ because the integrand
goes as $\rho^2$. From now on, throughout the paper, we will consider regularized free energies which are obtained as the difference of the $\rho^2$-divergent free energy $\CF$ with another $\rho^2$-divergent simpler term, namely
\begin{equation}
\label{eq:regular}
\Delta\CF = \CF -\int_{0}^{\infty} d\rho\hdue \rho^2
\end{equation}
in order to obtain a finite result. We will use this same regularization for every type of solution.

\subsection{Connected constant (blue) solutions}

Connected constant solutions have only an inter-layer condensate, meaning $f\neq 0$.
Because of the vanishing intra-layer condensate $c_1 =0$, as for the symmetric solution, we can simplify the expressions for $z'(\rho)$ and for the free energy by setting $l(\rho)=l'(\rho)=0.$ This gives
\begin{equation}
\label{eq:zprimecc}
z'(\rho)\huno=\huno \frac{2 f \rho^4}{\sqrt{(w_{h}^4 + \rho^4) \left(-f^2 \rho^8 + (w_{h}^4 - \rho^4)^2 (\rho^4 + q^2 \rho^4 + (w_{h}^4 + \rho^4)^2)\right)}}
\end{equation} 
and the separation is
\begin{equation}
\label{eq:lengthcc}
L\huno=\huno \int_{\rho_t}^{\infty}\huno d\rho\hdue \frac{2 f \rho^4}{\sqrt{(w_{h}^4 + \rho^4) \left(-f^2 \rho^8 + (w_{h}^4 - \rho^4)^2 (\rho^4 + q^2 \rho^4 + (w_{h}^4 + \rho^4)^2)\right)}}\, .
\end{equation}
The turning point $\rho_t$, where the denominator of the integrand vanishes, is given by
\begin{equation}
\label{eq:turncc}
\rho_t \huno=\huno \frac{1}{\sqrt{2}}\Bigl(-1-q^2 +\CJ + \sqrt{2}\sqrt{2f^2+ (1+q^2) \left(1+q^2 + 4w_{h}^4-\CJ \right)} \Bigr)^{1/4}
\end{equation}
where
$\CJ=\sqrt{4f^2 + (1+q^2 + 4w_{h}^4)^2}$.

The free energy for this solution is obtained by plugging $z'(\rho)$ as given in \eqref{eq:zprimecc} and $l=l'=0$
into the Routhian \eqref{eq:routh1} 
\begin{equation}
\label{eq:engcc}
\CF_{\mathrm{blue}}\huno=\huno \int_{\rho_t}^{\infty} \frac{w^4 - w_{h}^4}{w^6 \sqrt{w^4 + w_{h}^4}} \sqrt{\CH\huno\Bigl(1+\frac{f^2 w^{12}}{(w^4 - w_{h}^4)^2\hdue \CH-f^2 w^{12}} \Bigr)}
\end{equation}
where $\CH=q^2 w^8 + (w^8 + w_{h}^8 + w^4 + 2w^4 w_{h}^4) \rho^4$.

\subsection{Unconnected non-constant (red) solutions}

Unconnected non-constant solutions have a vanishing inter-layer condensate $f=0$ and a non-zero 
intra-layer condensate $c_1\neq 0$. The profile along the $z$ direction is flat, $z=\pm L/2$, while that 
along $l$ is not trivial and has to be determined by solving numerically the equation of motion.
The procedure to follow is quite involved. Let us explain it briefly.

First let us consider the charge neutral case, $q=0$.  In this case the solutions are Minkwoski embeddings
and they can be found by means of a shooting technique. On the one side we have the boundary condition at 
infinity \eqref{eq:asymptemb} with $c_0=0$, in terms of the modulus $c_1$. 
On the other side one can work out an expansion around $\rho=0$ for the solution where again 
a modulus, $l(0)$, appears. We can then ``shoot'' both from infinity and from zero varying the moduli 
and look for the solutions that smoothly intersect in an intermediate point $\rho=\rho_*$.

At finite charge density, as we already explained, only Black Hole embeddings are allowed.
In this case it seems not possible to use the same shooting procedure like before, 
since we cannot find a power series expansion for the solution for $l$ around the horizon. 
However it is still possible to set up a shooting procedure, where the shooting from the horizon is
done by simply imposing Neumann boundary condition.\footnote{Neumann boundary condition means that we have to fix the
	value of the derivative of $l$ at the horizon, which is the locus of points $(\rho_h,l(\rho_h))$ 
	in the first quadrant of the $(\rho,l)$-plane such that $\rho_h^2 + l(\rho_h)^2=w_h^2$.
	In principle we have no hints for the value of $l'(\rho_h)$ to be imposed at the horizon. 
	Because of the equation of motion we can only exclude that $l'(\rho_h)$ is zero (this can hold only for the trivial
	constant solution). However it turns out that the numerical integration, carried out with the 
	built-in Mathematica tool for numerical ODE solving, is insensitive to the value we impose for such a derivative.}

The free energy of these solutions is given by the Routhian \eqref{eq:routh2} with $f=0$ and $l(\rho)$ 
solution of equation of motion
\begin{equation}
\label{eq:engdiscnc}
\CF_{\mathrm{red}}\huno = \huno \int_{\rho_{\mathrm{min}}}^{\infty} d\rho\hdue\Bigl(\frac{w^4 - 
    w_{h}^4}{w^6}\Bigr)\huno\sqrt{\frac{\bigl( (q^2 w^8 + (w^4 + w^8 + 2 w^4 w_{h}^4 + w_{h}^8) \rho^4)         \bigr)\bigl(1+l'(\rho)^2 \bigr)}{w^4 +w_{h}^4}}\, .
\end{equation}
The turning point $\rho_{\mathrm{min}}$ is the point where the solution pinches off for a Minkowski embedding or 
a particular point in the event horizon, \ie such that $\rho_{\mathrm{min}}^2 + l(\rho_{\mathrm{min}})^2 = w_{h}^2$ 
for a Black Hole embedding.

\subsection{Connected non-constant (green) solutions}

The last kind of solutions is formed by the connected non-constant solutions where both $f$ and $c_1$ 
are non vanishing, meaning that both the inter- and intra-layer condensates are present. 

\begin{figure}[!ht]
\centerline{\includegraphics[scale=0.5]{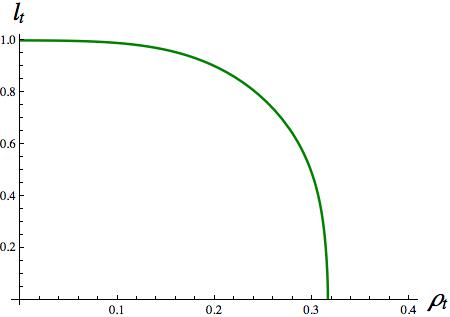}}
\caption{\small Plot of $l_t$ as a function of $\rho_t$ for $f=q=w_{h}=0.1$. A similar behavior holds for every value of the parameters.}
\label{fig:rhotmax}
\end{figure}

In order to find the solutions we have to solve a boundary value problem with two boundary conditions:
One is the usual asymptotic condition at $\rho\to \infty$, given by \eqref{eq:asymbeha}; 
the second one has to be given at the turning point $\rho=\rho_t$. Remember that
the brane and anti-brane worldvolumes have to join smoothly at $\rho_t$ and that in order to fulfill this requirement
we must have that $z'(\rho_t)=\infty$. Then the turning point is the point where
the denominator of \eqref{eq:finverse} is zero, namely
\begin{equation}\label{eq:turning}
	-f^2 w_t^{12}+\left(w_t^4 - w_{h}^4\right)^2 \left(q^2 w_t^8 +(w_t^4 +(w_t^4+w_{h}^4)^2)\rho_t^4\right)=0\, ,
\end{equation}
where we defined $l_t\equiv l(\rho_t)$ and $w_t\equiv \sqrt{\rho_t^2+l_t^2}$. This equation provides a relation between the turning point $\rho_t$
and the value of $l$ at the turning point, $l_t$. 
Plotting $l_t$ as a function of $\rho_t$ (see Figure~\ref{fig:rhotmax}) we find that there is a maximum turning point coordinate, call it $\rho_{t,\mathrm{max}}$.
Solving the eq.~\eqref{eq:turning} for $l_t$ we obtain $l_t$ as a function of the turning point value. We can then 
evaluate the equation of motion for $l$ at $\rho=\rho_t$, plugging the expression for $l_t$ from \eqref{eq:turning}. 
Doing this, the coefficient of $l''$ in the equation of motion vanishes and we are left with an expression 
which can be solved for $l_t '=l'(\rho_t)$, the derivative of the embedding function at the turning point.
In this way we basically obtain both the $l_t$ and $l_t '$ as a function of $\rho_t$. 
This information can be used to implement a shooting procedure to solve numerically the boundary value problem. When we shoot from 
infinity the parameter we vary is again $c_1$ while when we shoot from the turning point we vary $\rho_t$ itself.

Once we have the solution for $l$ we can compute the separation $L$ through \eqref{eq:separation} and the
free energy through the Routhian \eqref{eq:routh1} where $z'(\rho)$ is replaced by Eq.~\eqref{eq:finverse}
\begin{equation}
\label{eq:engcnc}
\CF_{\mathrm{green}}\huno=\huno \frac{(w^4 - w_{h}^4)^2 \huno \CH}{w^6}\sqrt{\frac{1+l'(\rho)^2}{(w^4 + w_{h}^4)\left(-f^2w^{12} + (w^4 - w_{h}^4)^2\hdue\CH\right)}}
\end{equation}
with $\CH=q^2 w^8 + (w^8 + w_{h}^8 + w^4 + 2w^4 w_{h}^4) \rho^4$.

\section{Free energy comparison}
\label{sec:free_en}

The main goal of the paper is to draw the full phase diagram for the D3/D5-\aD5-brane system. This is obtained by determining
the dominant configuration, \ie the one with least free energy, for each set of values of the variables considered, 
in our case the brane separation $L$, the charge density $q$ and the temperature $T\sim w_h$. 
We accomplish our goal by sectioning the variable space in lines of constant charge density 
and temperature. Basically we draw several plots of the free energy as a function of the separation for different fixed 
values of $q$ and $w_h$. This allows us to reconstruct the whole phase diagram.

\subsection{Neutral charge double monolayer}

\begin{figure}[!ht]
\centerline{\includegraphics[scale=0.4]{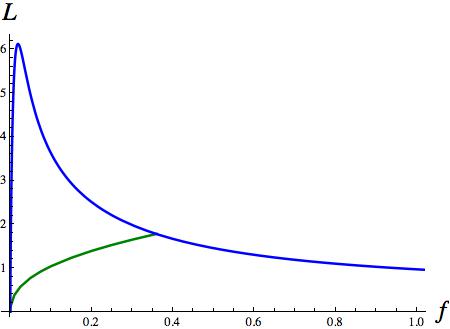}}
\caption{\small The separation of the monolayers, $L$, is plotted on the vertical axis and the parameter $f$ 
is plotted on the horizontal axis. The blue line is for the constant connected (blue) solution. 
The green line is for the connected $\rho$-dependent (green) solution. Here the temperature is $w_h =0.1$}
\label{fig:Lvsf}
\end{figure}

We start by considering the neutral charge case, $q=0$.
For the connected solutions, in order to draw the free energy as a function of the separation, we have to invert numerically the 
relation between $f$ and $L$. For such solutions the behavior of the brane separation as a function of the parameter $f$
is depicted in Figure~\ref{fig:Lvsf}. It is clear that for connected solutions, unlike in the zero temperature case, there exists a maximum value for the 
brane separation $L_\mathrm{max}$, which means that for separations $L>L_\mathrm{max}$ only disconnected solutions are allowed.

\begin{figure}[!ht]
\centerline{\includegraphics[scale=0.4]{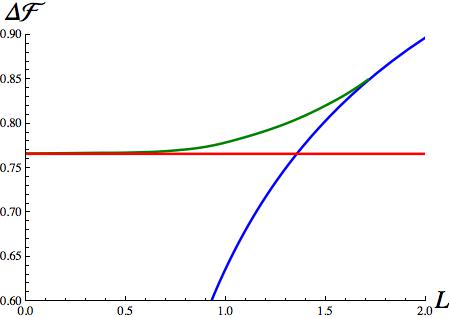}}
\caption{\small Neutral charge layers in a magnetic field. The regularized free energy  is plotted on the vertical axis, and the inter-layer separation $L$ is plotted on the horizontal axis. The blue line corresponds to the connected constant solution, the red line to the unconnected $\rho$-dependent solution, the black line to the symmetric solution and the green line to the connected $\rho$-dependent solution. The symmetric and green solutions never have the lowest energy. For large $L$, the red solution is preferred and for small $L$ the blue solution is more stable. }
\label{fig:EvsLneutral}
\end{figure}

Figure~\ref{fig:EvsLneutral} shows an example of plots of the free energies of all solutions as functions of $L$ 
for a temperature $w_h =0.1$.
This shows that increasing $L$ the system faces a first order transition from a  the connected constant solution with an inter-layer condensate only to the disconnected non-constant solution, with an intra-layer condensate only.
Even if we are now at finite temperature, we find a behavior analogous to the zero temperature one \cite{Evans:2013jma}.

\subsection{Charged monolayers}

\begin{figure}[!h]
	\begin{subfigure}{.49\textwidth}
		\centerline{\includegraphics[scale=0.4]{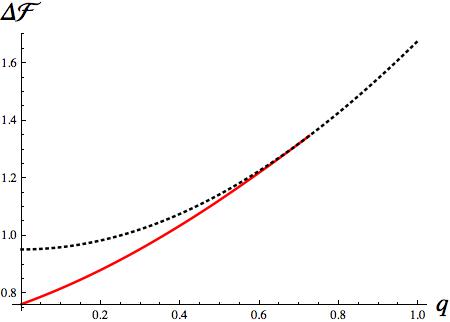}}
		\caption{\small $w_{h}=0.2$}
	\end{subfigure}
	\begin{subfigure}{.49\textwidth}
		\centerline{\includegraphics[scale=0.4]{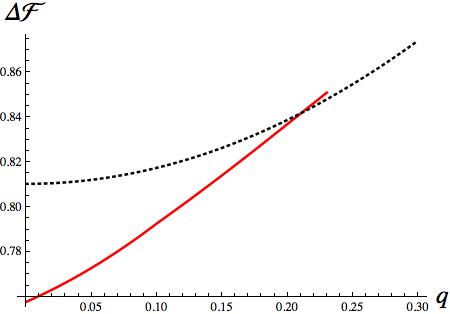}}
		\caption{\small $w_h =0.3$}
	\end{subfigure}
	\caption{\small The regularized free energy  is plotted on the vertical axis, and the charge density $q$ is plotted on the horizontal axis. The solid red line refers to the disconnected $\rho$-dependent solution and the dotted black line refers to the symmetric solution. For small charges the red phase is favorite. Then there is a phase transition after which the dominant phase is the chirally symmetric one. Note that we have different types of transition according to the value of the temperature as shown in \cite{Evans:2010hi}, where for $w_h >0.277$ they are first order and for $w_h <0.277$ they are second order phase transitions.}
	\label{fig:Ediscvsq}
\end{figure}

Before discussing the double monolayer system let us briefly review the monolayer case \cite{Evans:2010hi}.
This means that for the moment we consider only the disconnected solutions.
We know that raising both the temperature and the charge density leads to chiral symmetry restoration. In particular, for any given temperature (or charge density) there is a maximum value of the charge density (or temperature)  above which the stable solution is the symmetric one. Let us denote this value of the charge density as $q_\mathrm{crit}$, the critical charge density. As an example we show two of the plots of the free energies of the disconnected solutions as functions of $q$ in Figure~\ref{fig:Ediscvsq}.

\begin{figure}[!h]
	\centerline{\includegraphics[scale=0.375]{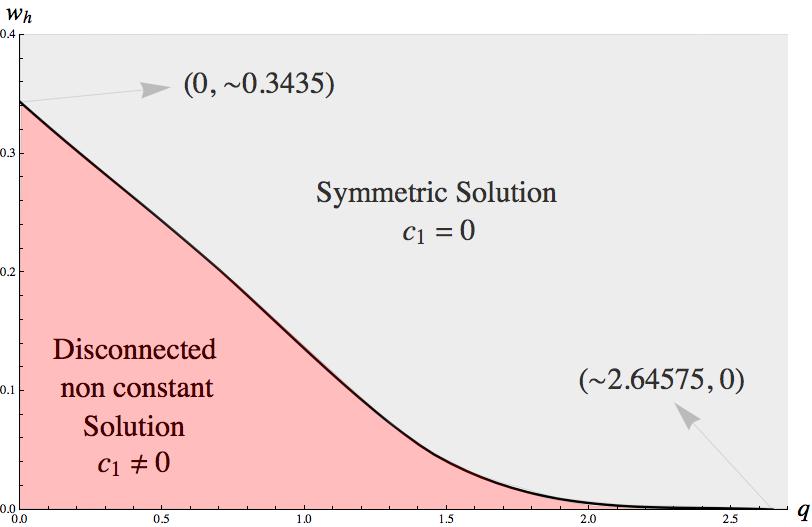}}
	\caption{\small The temperature is plotted on the vertical axis and the charge density $q$ is plotted on the horizontal axis. 
		We see that for large enough temperature or charge density the stable disconnected solution is the symmetric one.
		The maximum temperature for which one can still have a non-constant solution is $w_{h} \simeq 0.3435$ 
		(at $q=0$); the maximum charge is the critical charge at zero temperature, $q_\mathrm{crit}(w_h=0)=\sqrt{7}$. The latter point corresponds to a transition to a symmetric phase with BKT scaling.
		These results are in agreement with \cite{Evans:2010hi}.}
	\label{fig:Tvsqdiscphase}
\end{figure}

Note that as the temperature grows the corresponding critical charge density value gets smaller, till it vanishes for 
$w_{h} \simeq 0.3435$. For larger temperatures the D5-brane configurations with non constant profile are no longer allowed:
$w_{h} \simeq 0.3435$ is the maximum temperature for which the intra-layer condensate can form.
The single monolayer phase diagram is shown in Figure~\ref{fig:Tvsqdiscphase} \cite{Evans:2010hi}.

\subsection{Charge-balanced double monolayer}

Now we take into account also the connected configurations, in order to fully study the phase structure of the 
double monolayer system at finite temperature and finite density.

\begin{figure}[!ht]
\centerline{\includegraphics[scale=0.4]{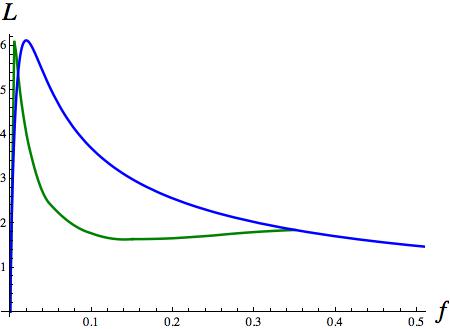}}
\caption{\small The separation of the monolayers, $L$, is plotted on the vertical axis and the parameter $f$ is plotted 
on the horizontal axis. The blue line is for the constant connected (blue) solution. 
The green line is for the connected $\rho$-dependent (green) solution.}
\label{fig:Lvsfcharge}
\end{figure}

Let us fix the value of the temperature and charge density to $w_h=0.1$ and $q=0.1$, respectively. 
The behavior of brane separation for the connected solutions as a function of $f$  is plotted in Figure~\ref{fig:Lvsfcharge}.
The behavior of $\rho$-dependent connected solution (green line) shows the presence of three branches: Starting from small $f$, first $L$ grows 
rapidly with $f$, till it reaches a maximum, $L_\mathrm{max}$; after that there is the decreasing branch, while finally, for large enough 
$f$, $L$ switches back to the growing behavior. 
When $q\to0$ the first two branches flatten on the vertical axis and eventually disappear for $q=0$,
so to recover the behavior of Figure~\ref{fig:Lvsf}.

\begin{figure}[!ht]
\begin{subfigure}{.49\textwidth}
\centerline{\includegraphics[scale=0.35]{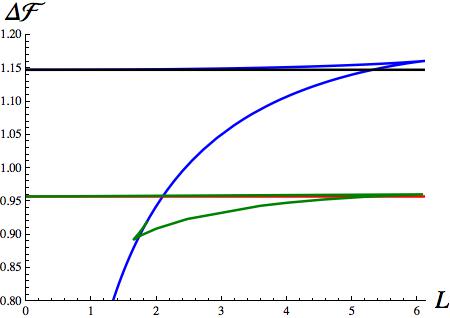}}
\caption{\small $q=0.25,\hdue w_h =0.1$}
\end{subfigure}
\begin{subfigure}{.49\textwidth}
\centerline{\includegraphics[scale=0.35]{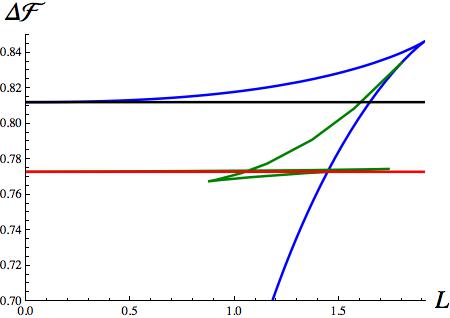}}
\caption{\small $q=0.05,\hdue w_h =0.3$}
\end{subfigure}
\caption{\small Charged layers in a magnetic field. The regularized free energy  is plotted on the vertical axis, and the inter-layer separation $L$ is plotted on the horizontal axis. The blue line corresponds to the connected constant solution, the red line to the unconnected $\rho$-dependent solution, the black line to the symmetric solution and the green line to the connected $\rho$-dependent solution. At small $L$ the dominant configuration is the blue connected constant with an inter-layer condensate. As the separation starts growing a first order phase transition occurs. Up on this transition length the green solution will remain the stable one with nonzero intra- and inter-layer condensates until it reaches a maximum separation where there is another first order phase transition in favor of a disconnected non constant red phase with only nonzero intra-layer condensate. }
\label{fig:EvsLcharge}
\end{figure}

Two plots of the free energies are shown in Figure~\ref{fig:EvsLcharge}.
These show that the dominant configuration is the connected constant one with an inter-layer condensate for small brane separation $L$.
Increasing $L$ the system first switches to the connected $\rho$-dependent phase, with both an intra-layer and an inter-layer condensate and
then, for larger separations, to the disconnected $\rho$-dependent.

\begin{figure}[!ht]
\begin{subfigure}{.48\textwidth}
\centerline{\includegraphics[scale=0.4]{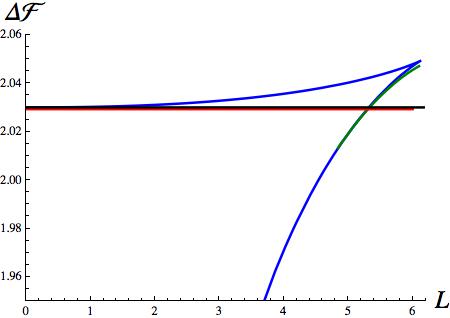}}
\caption{\small $q=1.1$}
\end{subfigure}
\begin{subfigure}{.48\textwidth}
\centerline{\includegraphics[scale=0.4]{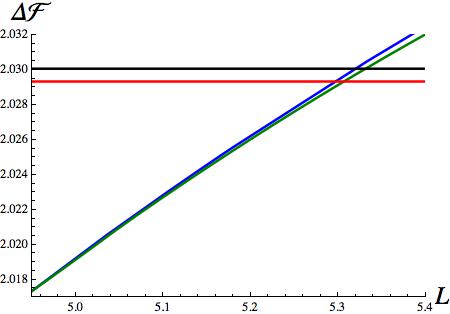}}
\caption{\small $q=1.1$ A closer look.}
\end{subfigure}

\begin{subfigure}{.48\textwidth}
\centerline{\includegraphics[scale=0.4]{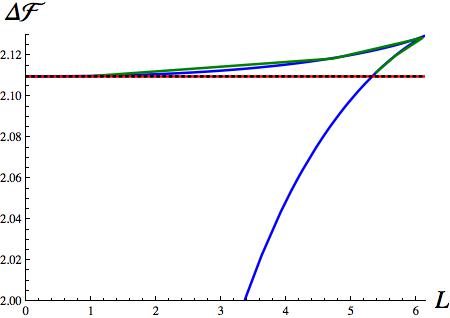}}
\caption{\small $q=1.15$}
\end{subfigure}
\begin{subfigure}{.48\textwidth}
\centerline{\includegraphics[scale=0.4]{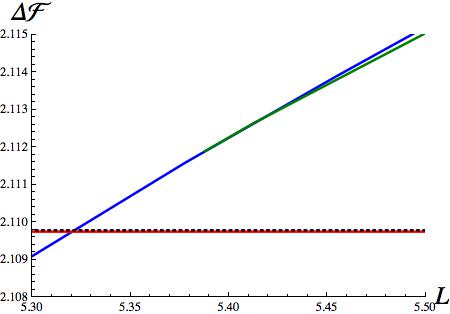}}
\caption{\small $q=1.15$ A closer look.}
\end{subfigure}
\caption{\small Plots of the regularized free energy as a function of the separation for different values of $q$. Note that in subfigure (d) the green solution already starts above the red solution and it is never the stable one. Here the only transition is between blue and red phases.}
\label{fig:EvsLqup}
\end{figure}

Now we analyze what happens if we increase the value of the charge density. It turns out that also the connected non constant solutions exist only when $q<q_\mathrm{crit}$, just like the disconnected non constant ones. 
Thus basically $q_{\mathrm{crit}}$ is the value of the charge above which only the trivial profile $l=0$ is allowed.
Let us fix the temperature to $w_h =0.1$. The critical charge density for such temperature is $q_{\mathrm{crit}}\simeq 1.163$. 
For fixed small charges $q<q_\mathrm{crit}$, increasing the separation the system undergoes two phase transitions. 
The first one is from the connected constant configuration, which is favored for small separations, to the connected non-constant 
configuration. Then for larger separation the disconnected non constant solution becomes dominant (see Figure~\ref{fig:EvsLcharge}). 
However this behavior is not valid for any $q<q_\mathrm{crit}$. For charge densities very close to $q_\mathrm{crit}$ indeed
there is a small domain $q^* < q < q_{\mathrm{crit}}$ where the green solution is never stable and by increasing $L$ the system faces only one transition,
between the connected constant configuration and the disconnected non-constant one (see Figure~\ref{fig:EvsLqup}).

\begin{figure}[!ht]
\begin{subfigure}{.49\textwidth}
\centerline{\includegraphics[scale=0.35]{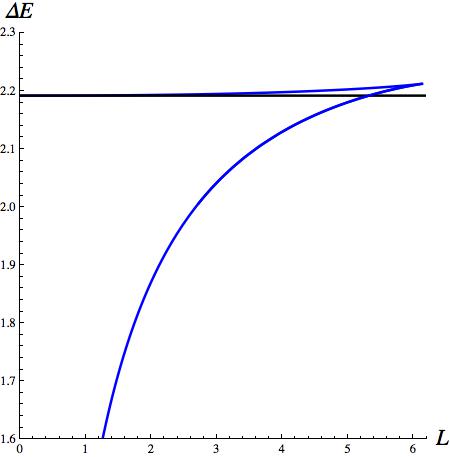}}
\caption{\small $q=1.2$}
\end{subfigure}
\begin{subfigure}{.49\textwidth}
\centerline{\includegraphics[scale=0.35]{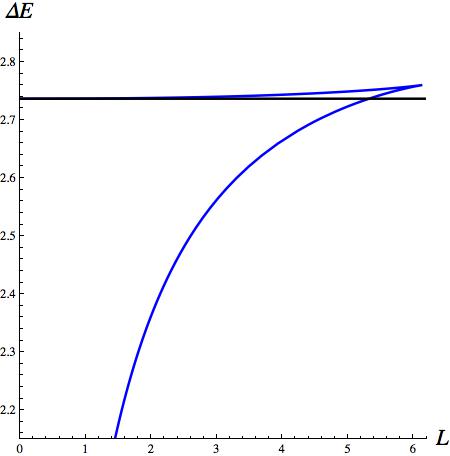}}
\caption{\small $q=1.5$}
\end{subfigure}
\caption{\small Plots of the regularized free energy as a function of the separation for $q>q_{\mathrm{crit}}.$ The blue line is the connected constant solution and the black one is the symmetric solution.}
\label{fig:EvsLconn}
\end{figure}
Once the charge density gets larger than $q_{\mathrm{crit}}$ the only allowed solutions are the constant ones. Two examples
of plots of the free energy as a function of the separation in such cases are shown in Figure~\ref{fig:EvsLconn}.
Note that the transition between blue and symmetric solutions takes up approximately the same separation for any value of the charge density 
(the slope of the corresponding transition curve in the phase diagram is nearly zero).

We performed the same computations for a wide range of temperatures $w_{h} < 0.3435$ and we found no
substantial modifications in the shapes of the curves of the energy as a function of the separation shown above. There
are only two differences: As one lowers the temperature the transition separation value $L_{\mathrm{max}}$
between the green and red phases increases and eventually reaches infinity in the limit $w_h \to 0$,
giving back the zero temperature phase diagram shown in \cite{Grignani:2014vaa}. The other difference is that
for large temperatures the $q_{\mathrm{crit}}$ and $q^∗$ points can be distinguished explicitly, while for small
temperatures they become almost coincident (see Figures~\ref{fig:phasediagram} and \ref{fig:phasediagram2}). In particular, in the
zero temperature limit, they both will take up the $q_{\mathrm{BKT}}=\sqrt{7}$ value in correspondence of
infinite separations.

\begin{figure}[!ht]
	\centerline{\includegraphics[scale=0.5]{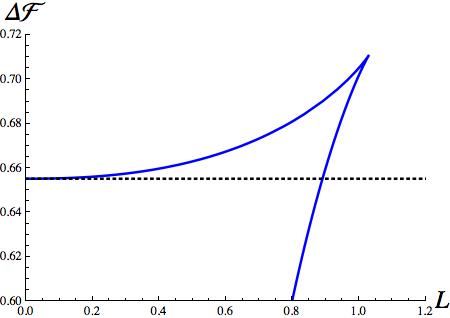}}
	\caption{\small Plots of the regularized free energy as a function of the separation for $q=0.5$ and $w_h=0.5$. The blue line is the connected constant solution and the black one is the symmetric solution.}
	\label{fig:EvsLhightemp}
\end{figure}

When $w_{h} > 0.3435$, no matter how small the charge is, the non constant solutions cease to exist and then the behavior is
analogous to the one we saw before for large charge densities, $q>q_{\mathrm{crit}}$.  
An example of the free energy comparison as a function of the separation is shown in Figure~\ref{fig:EvsLhightemp}.

\subsection{Phase diagram}

\begin{figure}[!ht]
\centerline{\includegraphics[scale=0.4]{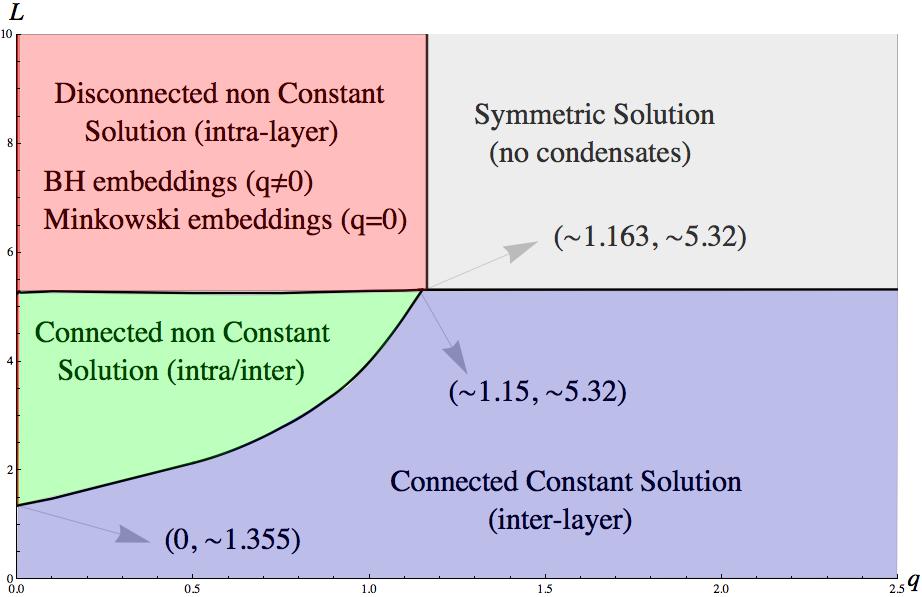}}
\caption{\small Phase diagram for $w_h = 0.1$: $L$ is plotted on the vertical axis and $q$ on the horizontal axis. The vertical black line corresponds to the critical value $q_{\mathrm{crit}}$ of the charge density at a given temperature that can be found in Figure~\ref{fig:Tvsqdiscphase}. The red line for neutral charge corresponds to a Minkowski embedding phase
and the red region to a black hole embedding phase; both have nonvanishing intra-layer condensate.
The blue region is the connected constant phase with only an inter-layer condensate. The green region is the connected non-constant phase with both nonvanishing intra- and inter-layer condensates. Finally the gray region is the chirally symmetric phase where both condensates are zero. Here $q^*\simeq 1.15$ and $q_{\mathrm{crit}}\simeq 1.163$}
\label{fig:phasediagram}
\end{figure}

\begin{figure}[!ht]
\centerline{\includegraphics[scale=0.4]{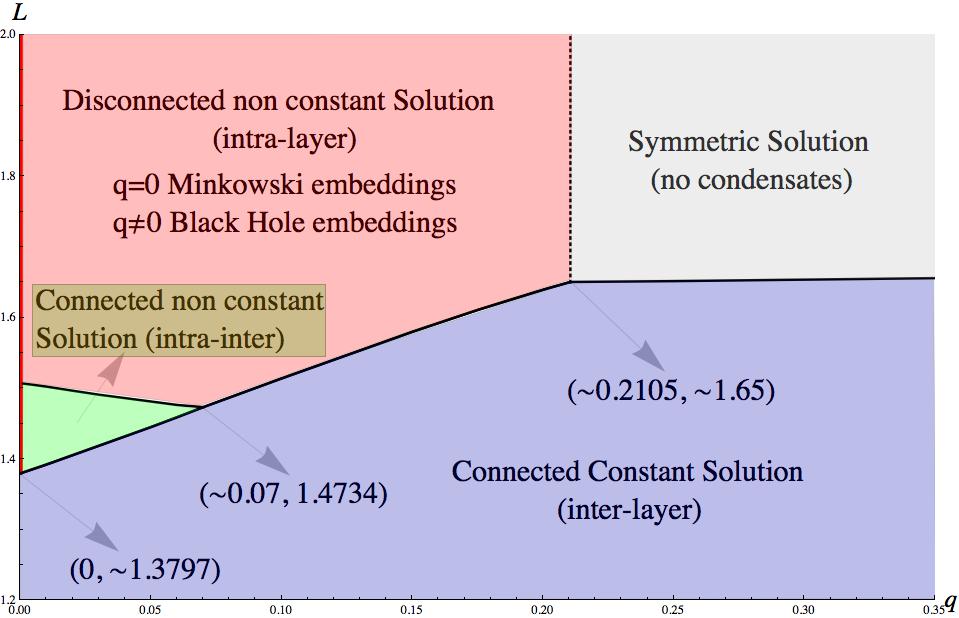}}
\caption{\small Phase diagram for $w_h = 0.3$. Here $q^*\simeq 0.07$ and $q_{\mathrm{crit}}\simeq 0.2105$.}
\label{fig:phasediagram2}
\end{figure}

\begin{figure}[!ht]
	\centerline{\includegraphics[scale=0.45]{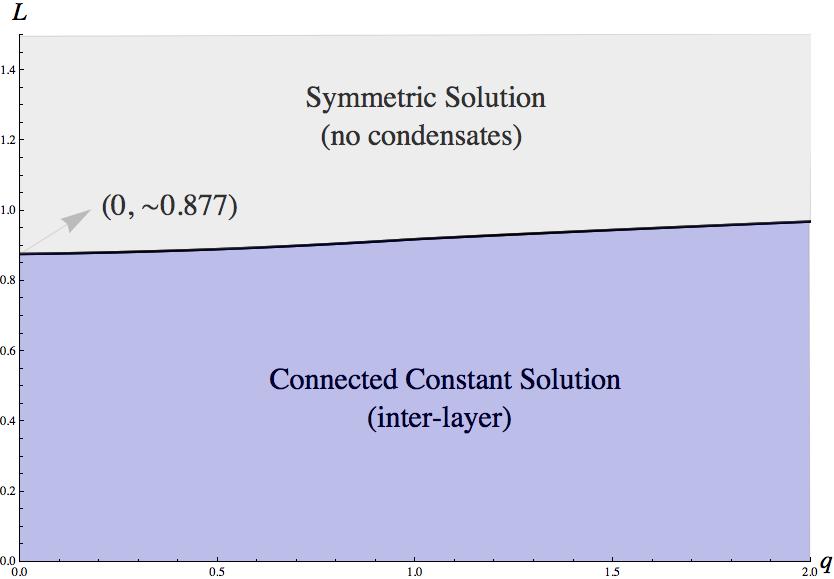}}
	\caption{\small Phase diagram for $w_h = 0.5$.}
	\label{fig:phasediagram3}
\end{figure}

The final step is to merge all the $(q,L)$ coordinates of the various transition points in some constant temperature phase diagrams. 
We show three examples of such phase diagrams for $w_h =0.1$, $w_h=0.3$ and $w_h=0.5$ in Figures~\ref{fig:phasediagram}-\ref{fig:phasediagram3}.


Let us briefly summarize the structure of the phase diagrams. First let us consider the case of temperatures $w_{h} < 0.3435$. For zero charge we have only two competitors: A connected constant blue phase with an inter-layer condensate $f\neq 0$ that is dominant for $L<L_{\mathrm{tr}}$ ($L_{\mathrm{tr}}\simeq 1.355$ for $w_h=0.1$ and $L_{\mathrm{tr}}\simeq 1.38$ for $w_h =0.3$) and a disconnected non-constant red double monolayer phase where the D5-branes are independent Minkowski embeddings with an intra-layer condensate for any $L>L_{\mathrm{tr}}$ separation. In the $0<q < q^*$ region, for small separations the blue solution is always favorite while for large separations the green connected non-constant phase with both condensates becomes the stable one. There is a first order phase transition between the two. In particular the green solution will be dominant for any value of $L>L_{\mathrm{tr}}$ until $L$ reaches  a maximum separation $L_{\mathrm{max}}$. Beyond $L_{\mathrm{max}}$ the stable phase is the disconnected $\rho$-dependent red one where the D5-branes have black hole embeddings. In the $q^* < q <q_{\mathrm{crit}}$ region the green solution, even if it still exists, is no longer stable and the only transition is between the blue and the red phases. Finally, for $q\geq q_{\mathrm{crit}}$ the green and red phases disappear and we find a competition just between the blue and symmetric configurations. The transition length between these two takes up approximately the same value for any value of the charge density larger that $q_{\mathrm{crit}}$. 

Note the different behavior of the transition lines between the green and red solutions for the $w_h =0.1$ and $w_h =0.3$ cases. 
In the $w_h =0.1$ phase diagram this line is almost horizontal, with a negligible positive slope $dL/dq >0$ while for the $w_h =0.3$ diagram it shows an explicit negative slope $dL/dq<0.$ The lowest charge we could analyze is $q=0.001$ in both cases: For lower values of $q$ the numerical analysis seems unreliable. 

Notice that as one raises the temperature, the symmetric solution domain has the tendency to enlarge and overcome the other non-symmetric solution domains.
This matches the expectation that high temperatures lead to chiral symmetry restoration.
For $w_{h} > 0.3435$, indeed the phase structure simplifies since the configurations with non-zero intra-layer condensate are no longer allowed. We are then left with only two relevant phases: The connected constant one with only the inter-layer condensate 
and the chirally symmetric one. The system switches from the former to the latter increasing the layer separation, as shown in the diagram in 
Figure~\ref{fig:phasediagram3}.

%



\section*{Acknowledgments}

We thank Nick Evans and Keun-Young Kim for useful discussions.

\appendix

\section{From natural units to cgs units}
\label{sec:cgsunits}

Throughout the paper, for ease of notation, we used natural units with the length scale $b^{-1/2}=\sqrt{\frac{\sqrt{\lambda}}{2\pi B}}$
set to one (by means of the rescalings \eqref{eq:replacements}), in such a way that all 
the physical quantities became adimensional. Now we are interested in re-convert the characteristic values we 
found for the physical quantities, like the separation, the magnetic field and the temperature,
in the more intelligible Gaussian cgs units, in order to make more apparent the connection with their real world material counterparts.

The rescaled magnetic field $b$ is related to the magnetic field in natural units $B_{\mathrm{nu}}$ through the relation 
\[
b=\frac{2\pi}{\sqrt{\lambda}}B_{\mathrm{nu}}.
\] 

Let us consider first the brane separation $L$. From \eqref{eq:replacements} we know that the values that appears in the phase diagrams
are actually the values of $L\rightarrow L\sqrt{b}$. Then we can write
\begin{equation}
\label{eq:Lcgs}
\bar{L}=L_{\mathrm{nu}} \sqrt{b}=L_{\mathrm{nu}} \frac{\sqrt{2\pi B_{\mathrm{nu}}}}{\lambda^{1/4}} \hspace{10pt} \rightarrow \hspace{10pt} L_{\mathrm{nu}}=\bar{L}\frac{\lambda^{1/4}}{\sqrt{2\pi B_{\mathrm{nu}}}}
\end{equation}
where $\bar{L}$ is the numerical value appearing on the vertical axis in the phase diagrams while 
$L_{\mathrm{nu}}$ is the separation expressed in natural units. 
We know that the magnetic field natural units are proportional to the inverse of a square length 
\[
[B_{\mathrm{nu}}]=\frac{1}{[\mathrm{length}]^2}\, ,
\] 
while a magnetic field in Gaussian units has 
\[
[B_{\mathrm{cgs}}]=\frac{[\mathrm{charge}]}{[\mathrm{length}]^2}\, .
\]
Since $\sqrt{\hbar c}$ has the unit of a charge we have that
\begin{equation}
\label{eq:Bcgs}
B_{\mathrm{nu}}=\frac{B_{\mathrm{cgs}}}{\sqrt{\hbar c}}\, .
\end{equation}
Plugging \eqref{eq:Bcgs} into eq.~\eqref{eq:Lcgs} we find
\begin{equation}
\label{eq:sepcgs}
L_{\mathrm{cgs}}=\bar{L} \hdue \frac{\lambda^{1/4}(\hbar c)^{1/4}}{\sqrt{2\pi B_{\mathrm{cgs}}}}
\end{equation}
where $L_{\mathrm{cgs}}$ is the separation in centimeters and $B_{\mathrm{cgs}}$ is the magnetic field expressed in Gauss. 
If we evaluate this expression for the typical values in play, like  $\bar{L}\simeq 1$ and \SI{1}{\tesla} magnetic field, 
$B_{\mathrm{cgs}}=10^4,$ we obtain
\begin{equation}
\label{eq:Lunits}
L\sim 3\times 10^{-7} \lambda^{1/4} \huno \si{\centi\metre}\, .
\end{equation}

For the temperature we can proceed in an analogous way. The temperature is related to
the horizon radius as follows
\[
(k_B T)_{\mathrm{nu}}=\frac{r_h}{\pi}=\frac{\sqrt{2} w_h}{\pi}
\]
and using \eqref{eq:replacements} we can write it as
\[
(k_B T)_{\mathrm{nu}}= \frac{\sqrt{2}\bar{w}_h \sqrt{2\pi B_{\mathrm{nu}}}}{\pi \lambda^{1/4}}\, ,
\]
where $\bar{w}_h$ is the numerical value as the temperature.
Now use again the relation \eqref{eq:Bcgs} for the magnetic field and we obtain
\begin{equation}
\label{eq:Tcgs}
T_{\mathrm{cgs}}=\frac{2\bar{w}_h(\hbar c)^{3/4} \sqrt{B_{\mathrm{cgs}}}}{k_B \sqrt{\pi} \lambda^{1/4}}
\end{equation}
where $T_{\mathrm{cgs}}$ is expressed in Kelvin.
Evaluating this expression for $1$ Tesla magnetic field and $\bar{w}_h=0.005$ yields
\begin{equation}
\label{eq:Tunits}
T\sim \frac{10^3}{\lambda^{1/4}} \hdue \si{\kelvin} \, .
\end{equation}

Note that the values we obtained for the layer separation and the temperatures are given in terms of an undetermined 
parameter, $\lambda$, which is the 't Hooft coupling, and it is proportional to the number 
of D3-branes and to the string coupling $g_s.$ Because the model we used describes a strongly coupled system, $\lambda$ 
has to be taken large. 

To give an idea of the orders of magnitude that can be found in room temperature superfluidity configurations, one can take the physical quantities to be for example $\lambda\sim 10\div 100$ in such a way that $\lambda^{1/4}\sim 1.78\div 3.16.$ Plugging it in \eqref{eq:Lunits} and \eqref{eq:Tunits} one finds $$ L\sim (5.3\div 9.5 )\times 10^{-7} \huno \si{\centi\metre}\, ,\hset\hset\hset\hset\hset T\sim (0.31\div 0.55)\times 10^3 \hdue \si{\kelvin}.$$




\bibliographystyle{newutphys}
\bibliography{biblio}
\addcontentsline{toc}{section}{\bf References}

\end{document}